\documentclass[multphys,vecphys]{svmult}

\usepackage{makeidx}         % allows index generation
\usepackage{graphicx}        % standard LaTeX graphics tool
\usepackage{multicol}        % used for the two-column index
\usepackage[bottom]{footmisc}% places footnotes at page bottom
\makeindex             % used for the subject index
\def\gtsim{\raise 2pt \hbox {$>$} \kern-1.1em \lower 4pt \hbox {$\sim$}}

\begin{document}

\title*{The Association of Compact Groups of Galaxies with Large-scale Structures}
\titlerunning{Compact Groups and Large-scale Structure}
\author{Heinz Andernach and Roger Coziol}
\authorrunning{H.\ Andernach \& R.\ Coziol}
\institute{Departamento de Astronom\'{\i}a, Universidad de
Guanajuato, Mexico\\
\texttt{heinz@astro.ugto.mx}, \texttt{rcoziol@astro.ugto.mx}}
\maketitle

\begin{abstract}
We use various samples of compact groups (CGs) to examine the types of
association CGs have with rich and poor clusters of galaxies at low ($z
\simeq 0.04$) and intermediate ($z\simeq 0.1$) redshifts. We find that
$\sim$10--20\% of CGs are associated with rich clusters and a much larger
fraction with poorer clusters or loose groups. Considering the incompleteness
of catalogs of poorer systems at intermediate redshift, our result is
consistent with {\it all} CGs at intermediate redshift being associated
with larger-scale systems.  The richness of the clusters associated
with CGs significantly increases from $z \simeq 0.04$ to $z\simeq 0.1$,
while their Bautz-Morgan type changes from early to late type for the
same range in $z$. Neither trend is compatible with a selection effect in the
cluster catalogs used.  We find earlier morphological types of galaxies
to be more frequent in CGs associated with larger-scale structures, compared
to those in CGs not associated to such structures.
We consider this as new evidence that CGs are part of the large-scale
structure formation process and that they may play an important role in
the evolution of galaxies in these structures.
\end{abstract}

\section{Motivation and Method}
\label{and1:sec2}
Although groups of galaxies form the principal
environment of most galaxies in the Universe, their origin and
evolution are still poorly understood. Theoretically, the
resolution of numerical simulations is still insufficient to
understand the formation of such small-scale structures.
Observationally, the intrinsic low richness of groups makes them
more difficult to detect and study than richer systems like
cluster of galaxies, especially at intermediate and high redshift.

Within the Hierarchical Structure Formation paradigm (based on CDM
models) we would like to address questions like e.g.\ ``When do groups
form and what IMF do they follow?'', or ``Assuming galaxy formation is 
biased, should we expect a sort of downsizing effect for groups, 
i.e.\ high-mass groups at high z forming before lower-mass ones at 
lower z?'' What is the role of groups in the formation of large-scale 
structures? If groups were
merging together to form larger-scale structures, how long can we
expect them to retain their dynamical characteristics and be
distinguished from the larger, more massive structure? Finally,
following the merging scenario, what is the importance of groups
in the evolution of galaxies located in large-scale structures?

Here we concentrate on Compact Groups of galaxies (CGs). The advantage 
of CGs over other groups is that they reach galaxy densities comparable 
to those of clusters, and are consequently easier to identify in the sky, 
even at moderate redshift.  From their relation with larger-scale structures
\cite{Mamon1989},\cite{RoodStruble1994},\cite{Ramellaetal1994},\cite{AndernachCoziol2005}
it seems clear also that CGs form part of the large-scale
structure formation process. Finally, it is expected that such
compact configurations should have some distinctive and
recognizable impact on the evolution of their galaxy members (see
e.g.\ Plauchu Frayn et al., these proceedings). A more detailed study of
the relation of CGs with large-scale structures may yield new insights 
on some of the above questions.

To examine the relation CGs have with large-scale structures 
we compiled a list of possible associations of CGs with rich and poor 
clusters of galaxies and with loose groups of galaxies. The CG samples 
we use are the HCGs \cite{hcg92}, the SCGs \cite{iovino2002}, the
SDSSCGs \cite{Leeetal2004}, and the PCGs \cite{deCarvalhoetal2005}. 
Note that in our previous analysis \cite{AndernachCoziol2005} only the 
preliminary PCG sample by \cite{iovinoetal2003} was available.
As samples of rich clusters we use the Abell catalog \cite{aco89},
and for poorer clusters and loose groups we take the NSC \cite{Galetal2003}
and USGC \cite{ramella2002} samples, respectively.
The association method we use is described in detail in
\cite{AndernachCoziol2005}.

Based on the position of the groups relative to the center of the
associated larger-scale structure and on a comparison of the
apparent magnitude of galaxies in both structures we identified
three types of such associations: ML or AML indicate that the galaxies 
in the CG are either the ``Most Luminous'' or ``Among the Most Luminous'' 
ones of the associated larger-scale structure, and SS if the group forms
some sort of substructure at the periphery of a larger system. 
Examples of each association class can be found in \cite{AndernachCoziol2005}.

\section{Results}
\label{and1:sec4}

\subsection{Associations of CGs with rich clusters}
\label{and1:sec4.1}

The associations of CGs with Abell clusters are described in 
Table~\ref{and1:tab1}. Columns are the CG sample name, total number of CGs in
each sample (note that on the basis of our optical inspection of all
459 published PCGs we discarded 52, mostly for stellar contamination),
the number of CGs associated with an Abell cluster, the percentage
of associated CGs, their mean redshift, the number distribution
of associated CGs among the different association types, and the
Abell richness classes as well as the Bautz-Morgan (BM) types of 
the associated clusters.  
Interesting differences are observed, which can easily
be related to the different group selection methods. Eye selection of
the CG candidates, like HCGs and visual inspection of the otherwise
automatically selected groups like the SDSSCGs, seem to discard most 
associations of CGs with larger-scale structures: 83\% of HCGs and 69\% of the SDSSCG are SS,
while 57\% are ML type in the SCGs and 72\% are AML in the PCGs.
This difference in the selection method may also explain why we
found only one coincidence of a SDSSCG with a PCG. Based on the 
surface density of SDSSCGs and PCGs and their region of overlap,
one would expect about 10 coincidences if the two selection 
methods were identical.

\begin{table}
\centering \caption{Associations of CGs with Abell clusters}
\label{and1:tab1}       % Give a unique label
\begin{tabular}{lcccccccccccccc}
\hline\noalign{\smallskip}
CG & N$_{\rm CG}^{\rm tot}$ & N$_{\rm ass}$ \ \ \ & \%\ \ &\ \ \ ~$\langle z\rangle$\ \ \ \ \ \
&\multicolumn{3}{c}{Ass. type}\ \ \
&\multicolumn{4}{c}{\ \ \ Richness\ }\ \ \
&\multicolumn{3}{c}{\ \ \ BM type \ }\\
& & & &  &SS&ML&AML~~&~~0&1&2&3~&~undef.&early&late \\
\noalign{\smallskip}\hline\noalign{\smallskip}
HCG      & 100 & ~6  & ~6\ & .04~& \,5&1   &... &5&... &... &1&... &2&3\\
SCG      & 121 & 21  & 17\ & .04~& \,9&12  &... &18&2&1&... &2&14&5\\
PCG      & 407 & 71  & 17\ & .12~& 20&... &51  &23&34&12&3&12&5&54\\
SDSSCG~~ & 177 & 16  & ~9\ & .12~& 11&... &5&  5&9&1&1&4&... &12\\
\noalign{\smallskip}\hline
\end{tabular}
\end{table}

Despite the difference in group selection techniques the CG-cluster
association rates seem quite similar at low ($z=0.04$) and
intermediate ($z=0.12$) redshifts.

Other differences observed are more difficult to explain. We note
for example that the richness of the associated clusters rises with
redshift: the fraction of R$=1$ cluster rises from  14\% in
the SCGs to 69\% in the PCGs and SDSSCGs. The average BM type
also seems to change with redshift: passing from 74\% early types (I,
I-II, II) in the SCGs to 92\% late types (II-III, III) in the PCG and
100\% late types in the SDSSCG.

We already observed these differences in our first examination of
these samples \cite{AndernachCoziol2005}. The fact that we
observe the same phenomenon using the complete (and revised)
sample of PCGs and the SDSSCG sample, confirms that these
differences occur in relation with the increase in redshift. In
\cite{AndernachCoziol2005} we verified that these differences
cannot be due to an incompleteness in richness of the
Abell/ACO sample at intermediate redshift, nor to some
sort of luminosity bias affecting clusters with different BM
types. These differences are consistent with an increase of the
mass of the associated structures at increasing redshift.

\subsection{Associations of CGs with poorer structures}
\label{and1:sec4.2}

The associations of CGs with large-scale structures poorer
than Abell clusters are described in Table~\ref{and1:tab2}.
Columns are the CG sample name, the sample name of the associated
structure, the number and percentage of CGs associated with that structure,
the mean redshift of the CG-associated structures, their distribution
among the three association types defined earlier, and the average
richness of the associated structure.

Comparing with Table~\ref{and1:tab1} we see that CGs are much more 
frequently associated with poorer than with richer structure.
This increase in the number of associations is significant: a factor 
of 2 to 6  for the SCG and HCG at low $z$, and a factor of 3 to 4 for the 
SDSSCG and PCG at intermediate $z$. Note that the severe incompleteness of
the NSC in low-richness structures (cf. \cite{Galetal2003}) implies that
at higher redshifts almost all CGs may be part of a larger-scale system.

In Table~\ref{and1:tab2} we also observe a significant increase in
the number of ML- and AML-type association which indicates that
CGs may generally form important ``substructures'' in these
poorer, larger-scale systems.

\begin{table}
\centering \caption{Associations of CGs with poorer structures}
\label{and1:tab2}       % Give a unique label
\begin{tabular}{lcccccccc}
\hline\noalign{\smallskip}
CG & LSS & N$_{\rm CG}^{\rm assoc}$\ \ \ & \%\ \ & \ $\langle z\rangle$\ \ \ \ \ &\multicolumn{3}{c}{Assoc. type} &~~$\langle {\rm N}_{\rm gal} \rangle$\\
   &        &  &       &                              & ~~~SS~ & ML & ~AML~~~  & \\
\noalign{\smallskip}\hline\noalign{\smallskip}
% &\multicolumn{7}{c}{loose groups (USGC)}\\
HCG      & USGC & 37&37&0.02~~&~~2&30&~5& 6\\
SCG      & USGC & 33&27&0.02~~&~~8&22&~3& 8\\
\hline\noalign{\smallskip}
% &\multicolumn{7}{c}{NSC clusters}\\
% \noalign{\smallskip}\hline\noalign{\smallskip}
PCG      & NSC & 195&62&0.13~~&~55&     10&130& 35\\
SDSSCG~~ & NSC &  21&25&0.13~~& ~~5&...& 14& 40\\
\noalign{\smallskip}\hline
\end{tabular}
\end{table}

\subsection{Variation of galaxy morphologies with CG environment}
\label{and1:sec4.3}

In Table~\ref{and1:tab3}, we compare the number of galaxies of early
(E-S0-S0/a), intermediate (Sa-Sab-Sb) and late (Sbc and later)
morphological type in CGs located in different large-scale environments. 
For both HCGs and SCGs, distinguishing whether they are or not associated
with clusters or loose groups (LGs), we list the number and percentages 
of galaxies of different morphological types.
N$_{\rm gal}$ is the number of galaxies for which we could find
a morphological type in NED ({\tt nedwww.ipac.caltech.edu}).
As expected, we find a significant increase of earlier morphological
types in CGs associated with large-scale structures. This result
suggests that in general the associations we found are physically
real.

\begin{table}
\centering \caption{Distribution of galaxy morphologies as a
function of CG environment}
\label{and1:tab3}       % Give a unique label
\begin{tabular}{lcccccccccc}
\hline\noalign{\smallskip}
 & \multicolumn{2}{c}{HCG} & \multicolumn{2}{c}{HCG} & \multicolumn{2}{c}{HCG}
 & \multicolumn{2}{c}{SCG} & \multicolumn{2}{c}{SCG} \\
Morphological~~~ & \multicolumn{2}{c}{~~out of cl.~~} & \multicolumn{2}{c}{~~~in LGs~~~} & \multicolumn{2}{c}{~~~in cl.~~~}
  & \multicolumn{2}{c}{~~~out of cl.~~~}& \multicolumn{2}{c}{~~in cl.~~}\\
% Morphological & \multicolumn{2}{c}{HCG out of Cl.} & \multicolumn{2}{c}{HCG in LG.} & \multicolumn{2}{c}{HCG in cl.}
%  & \multicolumn{2}{c}{SCG out of Cl.}& \multicolumn{2}{c}{SCG in Cl.}\\
type & N$_{\rm gal}$ &  \% & N$_{\rm gal}$ &  \% & N$_{\rm gal}$ &  \% & ~~~N$_{\rm gal}$ &  \% & N$_{\rm gal}$ &  \% \\
\noalign{\smallskip}\hline\noalign{\smallskip}
Early        & 142 &  54.4 & ~~73& 46.8 &~~17& 65.4 & ~~~~72&45.0 &~~22&64.7\\
Intermediate~~~~ &  33 &  12.6 & ~~28& 17.9 &~~~2&  7.7 & ~~~~49&30.6 &~~~7&20.6\\
Late         &  86 &  33.0 & ~~55& 35.3 &~~~7& 26.9 & ~~~~39&24.4 &~~~5&14.7\\
\hline\noalign{\smallskip}
\end{tabular}
\end{table}

On the other hand, it is also quite interesting to note how rich
in early-type galaxies the isolated CGs already are.  Note that
there seems to be no difference between isolated CGs and those
associated with loose groups (LG). These results suggest that
compact configurations like CGs have a great influence on galaxy 
evolution. This may be independent of the environment or it may
depend on some threshold mass of the larger-scale structure with 
which the CG is associated.

\section{Conclusions}
\label{and1:sec5}

Despite the obvious biases introduced by the different selection
methods, the high number of associations of CGs with larger-scale
structures encountered in our analysis indicates that these
systems must form naturally during the large-scale structure
formation process. Both, the increase of mass of the associated
structures at intermediate redshift, and the variation of
galaxy morphologies with CG environment, are consistent with 
biased galaxy formation, which implies some sort of downsizing 
effect for groups: more massive groups form before less massive ones. 
Compact groups may affect galaxy evolution independently of
environment, and/or the CG environment may affect galaxies more
strongly above some threshold mass of the structures in which the
groups form.

% BibTeX users please use
% \bibliographystyle{}
% \bibliography{}
%
% Non-BibTeX users please follow the syntax
% the syntax of "referenc.tex" for your own citations
%%%%%%%%%%%%%%%%%%%%%%%% referenc.tex %%%%%%%%%%%%%%%%%%%%%%%%%%%%%%
% sample references
% "physics"
%
% Use this file as a template for your own input.
%
%%%%%%%%%%%%%%%%%%%%%%%% Springer-Verlag %%%%%%%%%%%%%%%%%%%%%%%%%%

%
% BibTeX users please use
% \bibliographystyle{}
% \bibliography{}

%%%%%%%%%%%%%%%%%%%%%%%%%%%%%%%%%%%%%%%%%%%%%%%%%%%%%%%%%%%%%%%%%%%%%%  }

%%%%%%%%%%%%%%%%%%%%%%%%%%%%%%%%%%%%%%%%%%%%%%%%%%%%%%%%%%%%%%%%%%%%%%

\printindex
\end{document}